\documentclass[prc,floatfix,groupedaddress,nofootinbib,showpacs,preprintnumbers,amsmath,amssymb,amsfonts,superscriptaddress,widetable]{revtex4-1}
\usepackage{bm}
\usepackage{mathrsfs}
\usepackage{amssymb}
\usepackage{amsmath}
\usepackage{graphicx}
\usepackage{array}
\usepackage{makecell}
\usepackage[usenames,x11names]{xcolor}

\begin{document}
\title{Impact of the equation-of-state -- gravity degeneracy on constraining the nuclear symmetry energy from astrophysical observables}
\author{Xiao-Tao He}\email{hext@nuaa.edu.cn}
\affiliation{Department of Physics and Astronomy, Texas A\&M
               University-Commerce, Commerce, TX 75429, USA}
\affiliation{College of Material Science and Technology, Nanjing
University of Aeronautics and Astronautics, Nanjing 210016, China}
\author{F.~J. Fattoyev}\email{farrooh.fattoyev@tamuc.edu}
\affiliation{Department of Physics and Astronomy, Texas A\&M
               University-Commerce, Commerce, TX 75429, USA}
\author{Bao-An Li}\email{bao-an.li@tamuc.edu}
\affiliation{Department of Physics and Astronomy, Texas A\&M
               University-Commerce, Commerce, TX 75429, USA}
\author{W.~G. Newton}\email{william.newton@tamuc.edu}
\affiliation{Department of Physics and Astronomy, Texas A\&M
               University-Commerce, Commerce, TX 75429, USA}
\date{\today}
\begin{abstract}
There is a degeneracy between the equation of state (EOS) of
super-dense neutron-rich nuclear matter and the strong-field gravity
in understanding properties of neutron stars. While the EOS is still
poorly known, there are also longstanding ambiguities in choosing
Einstein's General Relativity (GR) or alternative gravity theories
in the not-so-well tested strong-field regime. Besides possible
appearance of hyperons and new phases, the most uncertain part of
the nucleonic EOS is currently the density dependence of nuclear
symmetry energy especially at supra-saturation densities. At the
same time, the EOS of symmetric nuclear matter (SNM) has been
significantly constrained at saturation and supra-saturation
densities. To provide information that may help break the
EOS-gravity degeneracy, we investigate effects of nuclear symmetry
energy within its uncertain range determined by recent terrestrial
nuclear laboratory experiments on the gravitational binding energy
and space-time curvature of neutron stars within GR and the
scalar-tensor subset of alternative gravity models, constrained by
recent measurements of the relativistic binary pulsars J1738+0333
and J0348+0432. In particular, we focus on effects of the following
three parameters characterizing the EOS of super-dense neutron-rich
nucleonic matter: (1) the incompressibility $K_0$ of symmetric
nuclear matter (SNM), (2) the slope $L$ of nuclear symmetry energy
at saturation density and (3) the high-density behavior of nuclear
symmetry energy. We find that the variation of either the density
slope $L$ or the high-density behavior of nuclear symmetry energy
leads to large changes in both the binding energy and curvature of
neutron stars while effects of varying the more constrained $K_0$
are negligibly small. The difference in predictions using the GR and
the scalar-tensor theory appears only for massive neutron stars, and
even then is significantly smaller than the differences resulting
from variations in the symmetry energy. We conclude that within the
scalar-tensor subset of gravity models, the EOS-gravity degeneracy
has been broken by the recent relativistic pulsar measurements, and
that measurements of neutron star properties sensitive to the
compactness constrain mainly the density dependence of the symmetry
energy at saturation and supra-saturation densities.

\end{abstract}

\smallskip

\maketitle

\section{Introduction}
\label{sec:intro}

Neutron stars (NSs) are among the most dense and
neutron-rich objects in the Universe. They are thus a natural
testing ground for fundamental theories of strong-field gravity and
the nature of matter under extreme conditions. Our knowledge about
super-dense matter in NSs remains quite poor, see, e.g.,
~\cite{Lattimer:2000nx,Lattimer:2006xb,Lattimer:2012nd}, and may be
compounded by the possible break-down of Einstein's General
Relativity (GR) in the strong-field
regime~\cite{Pani:2011xm,Eksi:2014wia}. There is a possible degeneracy
between the nuclear equation of state (EOS) for matter in NSs and
the theory of gravity applied to describe their structure. How to
break this degeneracy is a longstanding problem to which many recent
studies have been devoted: see, e.g.,
~\cite{Will:2005va,Harada:1998ge,Sotani:2004rq,Lasky:2008fs,Wen:2009av,
Cooney:2009rr,Horbatsch:2010hj,Arapoglu:2010rz,Deliduman:2011nw,
Pani:2011xm,Sotani:2012aa,Xu:2012wc,Lin:2013rea,Yagi:2013mbt,Sotani:2014goa,
Eksi:2014wia}.  A better understanding of the EOS will certainly
help constrain the strong-field gravity theory, and vice versa.
Besides the possible appearance of hyperons and quarks in
super-dense matter in NSs, the most uncertain part of the nucleonic
EOS is currently the nuclear symmetry energy at supra-saturation
densities~\cite{Li:2008gp}. Ideally, effects of the uncertain nuclear EOS should be examined
simultaneously within GR and a large sample of alternative gravity
theories. This however requires much collaborative work of the whole
community. In this work, we examine the extent to which a
gravity-EOS degeneracy exists when models of gravity are restricted
to GR and Scalar-Tensor (ST) theories, and when variations of the
EOS are the result of variation of (1) the incompressibility $K_0$
of symmetric nuclear matter (SNM), (2) the slope $L$ of nuclear
symmetry energy  at saturation density and (3) the high-density
behavior of nuclear symmetry energy, within ranges constrained by
terrestrial nuclear data. We find that the variation of either the
density slope $L$ or the high-density behavior of nuclear symmetry
energy within their uncertainty ranges lead to significant changes
in both the binding energy and curvature of NSs, while effects of
varying $K_0$ are negligibly small. In particular, the
variations are significantly greater than those that result from ST
theories of gravity, leading to the conclusion that within those
subset of gravity models, measurements of neutron star properties
constrain mainly the EOS.

Neutron stars are very compact: the typical surface compactness parameter
of NSs is of the order of $\eta_{R} \equiv 2GM/c^2R \approx 0.4$.
While it is common to measure the strength of the gravitational
field with the compactness parameter, this leaves out most of the
useful information related to its strength, in particular, when one
would like to test the deviations of GR from alternative theories of
gravity in the strong-field regime. It was shown in
Ref.~\cite{Psaltis:2008bb} that for the gravitational field to be
considered as {\sl strong} it is more natural to choose a parameter
$\xi \equiv 2GM/c^2r^3$, which is related to the non-vanishing
components of the Riemann tensor in vacuum, ${\mathcal{R}^1}_{010} =
- \xi$. Notice that the curvature is the lowest order quantity of
the gravitational field that cannot be set to zero by a coordinate
transformation~\cite{Psaltis:2008bb}. Recently, by using the
venerable Akmal-Pandharipande EOS~\cite{Akmal:1997ft} the authors of
Ref.~\cite{Eksi:2014wia} have studied the radial profile of
compactness and curvature in neutron stars and have concluded that
the GR is in a less tested regime than the EOS over the entire
region of neutron star. In other work, by using a very stiff EOS
with nuclear incompressibility of $K_{0}=546$
MeV~\cite{Haensel:1981aa}, it was shown that the fractional
gravitational binding energy of the PSR J0348+0432 with mass
$2.01\pm 0.04M_{\odot}$~\cite{Antoniadis:2013pzd} is twice as large
as that of both NSs in the celebrated double pulsar system PSR
J0737-3039, whose masses have been measured with a very high
precision. Given that the observation of these pulsars resulted in
some of the most precise tests of GR to date, it was concluded that
the great difference in their fractional gravitational binding
energies keeps PSR J0348+0432 in a special place far outside the
presently tested binding energy range, and therefore this NS could
serve as a testbed for theories of strong-field gravity.
Particularly interesting is the study of the scalar-tensor theory of
gravity, whose parameters have a well-defined constraint from a
ten-year pulsar timing observation of PSR
J1738+0333~\cite{Freire:2012mg}.

Disentangling the EOS-Gravity degeneracy is important if we are to
make use of astrophysical observables to measure neutron
star properties. Potential observables directly related to the neutron
star compactness include the binding energy \cite{Podsiadlowski:2005ig,Newton:2009vz},
surface redshift \cite{Majczyna:2006tv} and maximum spin frequency of a neutron star \cite{Haensel:1989}, and it is these observables that we shall
focus on in this paper.

The nuclear symmetry energy $S(\rho)$ represents the penalty imposed
on the energy of the nuclear system as one departs from the
symmetric limit of equal number of protons and neutrons at density
$\rho$. Within the parabolic approximation of the nuclear EOS for
neutron-rich matter, $S(\rho)$ is simply the difference in specific
energy between pure neutron matter (PNM) and SNM. The bulk
parameters characterizing the EOS of SNM near nuclear saturation
density $\rho_0$ are well constrained from fitting models to the
ground state properties of finite nuclei, and to the {\sl
breathing-mode} energies of giant resonances in medium-to-heavy
nuclei~\cite{Youngblood:1999,Agrawal:2003xb,Colo:2004mj,Garg:2006vc,
Shlomo:2006aa, Piekarewicz:2009gb, Chen:2011ps, Patel:2012zd}. The incompressibility of nuclear
matter is constrained in the far tighter range of $K_0 = 240 \pm 20$ MeV, which
is well below the value predicted three decades ago, for example, in
Ref.~\cite{Haensel:1981aa}. Moreover, combining information from
studying the collective flow and kaon production in relativistic
heavy-ion collisions in several terrestrial nuclear physics
laboratories the EOS of SNM at higher densities has also been
limited in a relatively small range up to about
$4.5\rho_0$~\cite{Danielewicz:2002pu} consistent with the latest
observations of two-solar mass neutron
stars~\cite{Demorest:2010bx,Antoniadis:2013pzd}. However, despite
intensive efforts devoted to constraining the density dependence of
the nuclear symmetry energy, its knowledge still remains largely
uncertain even around nuclear saturation density
$\rho_0$~\cite{Chen:2004si, Shetty:2007zg, Klimkiewicz:2007zz,
Tsang:2008fd, Centelles:2008vu, Xu:2010fh, Steiner:2011ft,
Lattimer:2012xj, Tsang:2012se, Fattoyev:2012uu, Li:2013ola,
Newton:2011dw, Fattoyev:2013yaa}. It plays a vital role
not only in describing the structure of rare isotopes and their
reaction
mechanisms~\cite{Li:1997px,Baran:2004ih,Chen:2007fsa,Li:2008gp}, but
also determines uniquely the proton fraction essential for
understanding the cooling mechanism and appearance of exotic species
in neutron stars~\cite{Lattimer:2006xb,Steiner:2004fi}. Moreover, it
affects significantly the structure, such as the radii, moment of
inertia, tidal polarizability and the core-crust transition density,
as well as the frequencies and damping times of various oscillation
modes of neutron stars (For a full review please refer to the
topical issue on this subject matter~\cite{Li:2014nse}).
Furthermore, the uncertainty in the high-density behavior of the
symmetry energy is quite large with predictions from all varieties
of nuclear models diverging dramatically~\cite{Brown:2000pd,
Szmaglinski:2006fz, Dieperink:2003vs}. Besides the well known
difficulties of treating accurately quantum many-body problems,  our
poor knowledge about the spin-isospin dependence of three-body and
many-body forces, the short-range behavior of nuclear tensor force
and the resulting isospin-dependence of short-range nucloen-nucleon
correlations, coupled with the  lack of sensitive experimental
probes are mainly responsible for the current situation. Whereas
several observables have been proposed in the past~\cite{Li:2008gp},
and some hints of the high-density behavior of nuclear symmetry
energy have been reported recently~\cite{Xiao:2009zza,
Russotto:2011hq}, there is still no widely accepted conclusion
regarding the density dependence of nuclear symmetry
energy~\cite{Trautmann:2012nk}. Therefore, the symmetry energy is
still regarded as the most uncertain part of the EOS of neutron-rich
nuclear matter.

The paper has been organized as follows. First, in
Sec.~\ref{sec:EOS} we will overview the underlying nuclear EOS
models used in this study. Special attention is paid to the
uncertainties of the EOS of both SNM at saturation and the nuclear
symmetry energy at both saturation and supra-saturation densities.
Second, in Sec.~\ref{sec:BE} we provide the necessary details
required to calculate the gravitational binding energy and discuss
its sensitivity to the nuclear symmetry energy, in particular. Next,
in Sec.~\ref{sec:Curv} we discuss the role of the full contraction
of Riemann curvature tensor as a suitable choice of curvature. In
particular, we will show that while the relative strength of the
gravitational field could be several times larger from the surface
to the core of the neutron star, it strongly depends on the choice
of the underlying EOS. Then, in Sec.~\ref{sec:ScalarTensor} we will
briefly overview the necessary formula to calculate the mass versus
radius relation and the gravitational binding energy in the
scalar-tensor models of gravity. We will show that current
observational constraints on parameters of this theory puts them at
a disadvantage in breaking the degeneracy between the uncertainties
in the EOS and alternative models of gravity.
Later, in Sec.~\ref{sec:AstrophObserv} we discuss the possibility of
determination of the nuclear symmetry energy from two particular astrophysical
observables sensitive to the compactness of the star: the surface gravitational redshift of non-rotating NSs
and the maximum spin frequency of rapidly rotating NSs.
And finally in Sec.~\ref{sec:Conclusion} we offer our concluding
remarks.

\section{The Equation of State of Stellar Matter}
\label{sec:EOS}

The equation of state of neutron-rich nucleonic matter, in general,
can be expressed as
\begin{eqnarray}
E(\rho,\alpha)=E_{0}(\rho)+\mathcal{S}(\rho)\alpha^{2}+\mathcal{O}(\alpha^{4})
\ , \label{eq:E}
\end{eqnarray}
where $E(\rho,\alpha)$ and $E_0(\rho)$ are the binding energy per
nucleon in asymmetric nuclear matter of isospin asymmetry
$\alpha=(\rho_n-\rho_p)/\rho$ and in SNM ($\alpha=0$), respectively.
Here $\mathcal{S}(\rho)$ is referred to as the nuclear symmetry
energy, which represents the energy cost of departing from symmetric
limit of equal neutrons and protons, and $\rho=\rho_{n}+\rho_{p}$ is
the total baryon density with $\rho_{n}$ ($\rho_{p}$) being the
neutron (proton) density. It is customary to characterize the EOS
further in terms of bulk nuclear parameters by expanding both
$E_{0}(\rho)$ and $\mathcal{S}(\rho)$ in Taylor series around
nuclear saturation density
\begin{eqnarray}
E_{0}(\rho)&=&B_0 +\frac{1}{2} K_{0} \chi^{2}+\mathcal{O}(\chi^{3}) \ , \\
\mathcal{S}(\rho)&=&J+L\chi+\frac{1}{2} K_{\rm sym}
\chi^{2}+\mathcal{O}(\chi^{3}) \ ,
\end{eqnarray}
where $\chi \equiv (\rho-\rho_{0})/3\rho_{0}$ quantifies the
deviations of the density from its saturation value. As discussed in
the Introduction the binding energy at saturation $B_0$ and
the nuclear incompressibility coefficient $K_{0}$ are tightly
constrained by terrestrial experimental data using ground state
properties of finite nuclei and energies of giant resonances. On the
other hand, while the symmetry energy at saturation $J$ is more or
less known, its density slope $L$ is largely unconstrained. Moreover
the value of the curvature of the symmetry energy at saturation
$K_{\rm sym}$ still has a huge uncertain range.
\begin{figure}[h]
\smallskip
\includegraphics[width=5in,angle=0]{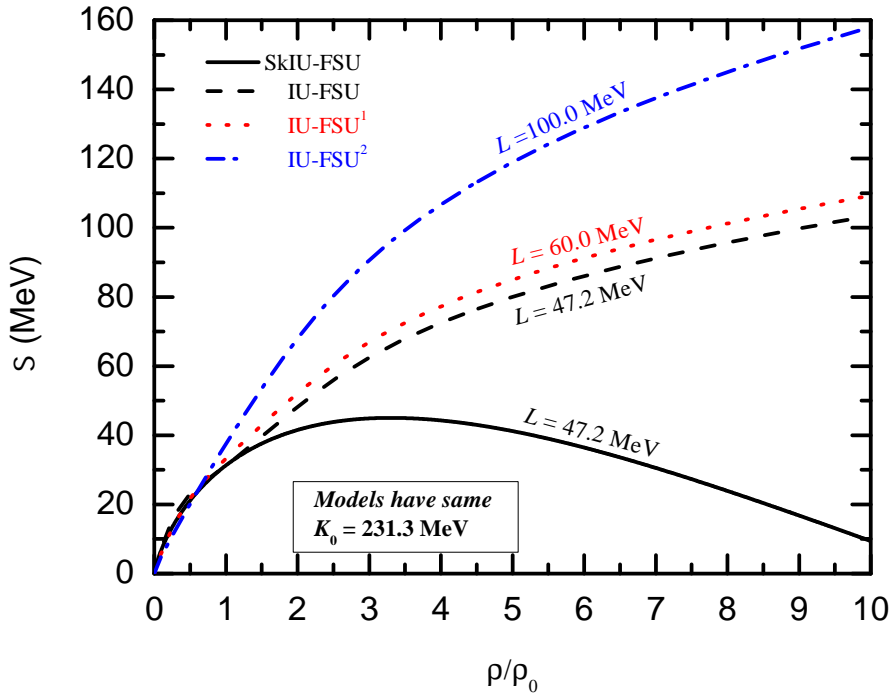}
\caption{(color online). Density dependence of the nuclear symmetry
energy for the four models discussed in the text. The solid black
line corresponds to the symmetry energy predictions in SkIU-FSU, the
dashed black line corresponds to the original IU-FSU model, the dotted
red and dash-dotted blue lines are symmetry energy predictions in
the IU-FSU-like models with density slope $L=60.0$ MeV (IU-FSU$^1$)
and $L=100.0$ MeV (IU-FSU$^2$), respectively. Notice that we did not
display results for the IU-FSU-like models with incompressibility
coefficients of $K_0 = 220$ MeV and $K_0 = 260$ MeV, since their
predictions are almost identical to the one of the original IU-FSU
model.} \label{Fig:EOS}
\end{figure}

As our base models we employ two recently established EOSs for
neutron-rich nucleonic matter within the IU-FSU Relativistic Mean
Field (RMF) model and the SkIU-FSU Skyrme-Hartree-Fork (SHF)
approach~\cite{Fattoyev:2012ch,Fattoyev:2010mx,Fattoyev:2012uu}. The
IU-FSU EOS was obtained by adjusting the parameters of the RMF to
satisfy the latest constraints from terrestrial nuclear experiments
and astrophysical observations~\cite{Fattoyev:2010mx}. Moreover, its
predictions closely match the state-of-the-art nuclear many-body
calculations for the EOS of low-density pure neutron matter (PNM).
By construction, the SkIU-FSU model is a non-relativistic
counterpart of the IU-FSU, and they have similar EOSs for both SNM
with $K_0=231.3$ MeV, and PNM around and below $\rho_0$. In
particular, the slope of the symmetry energy at saturation for both
models is $L = 47.2$ MeV. At subsaturation densities,
$\mathcal{S}(\rho)$ is almost identical for these models. However,
their predictions for the symmetry energy at super-saturation
densities of $\gtrsim 1.5\rho_0$ significantly differ with the
SkIU-FSU leading to a much softer $\mathcal{S}(\rho)$ at higher
densities (See Fig. \ref{Fig:EOS}). For full details on these models
we refer the reader to
Refs.~\cite{Fattoyev:2012ch,Fattoyev:2010mx,Fattoyev:2012uu}.

To test the sensitivity of the gravitational binding energy and
space-time curvature of neutron stars to the variations of
properties of neutron-rich nuclear matter around saturation density,
we have further introduced four additional EOSs using the IU-FSU as
the base model. By tuning two purely isovector parameters of the
IU-FSU model one can efficiently modify the density dependence of
the symmetry energy~\cite{Fattoyev:2012ch}. This tuning is carried
out under the constraint that the value of the symmetry energy at a
subsaturation density of $\rho \approx 0.103 \rho_0$ is fixed at
$26.0$ MeV. Such tuning leaves the EOS of SNM unchanged and
guarantees that predictions for the ground-state properties of
finite nuclei will not deviate from their experimental
values~\cite{Horowitz:2000xj}. In this way we created two RMF models
with density slopes of the symmetry energy at saturation density of
$L=60$ MeV and $L=100$ MeV. Finally, by properly adjusting all
parameters of the RMF model, so that their bulk properties match
those of the IU-FSU baseline model, we have created two additional
RMF models with differing incompressibility coefficients of $K_0 =
220$ MeV and $K_0 = 260$ MeV. We have ensured that predictions of
these two models reproduce the experimental data on charge radii and
binding energies of some selected closed-shell nuclei by slightly
adjusting the binding energy at saturation. Shown in
Fig.~\ref{Fig:EOS} are the density dependence of the symmetry energy
for four of these models. Notice we did not display models with $K_0
= 220$ MeV and $K_0 = 260$ MeV, because the density dependence of
the symmetry energy in these models are almost indistinguishable
from that of the original IU-FSU.

\section{Gravitational Binding Energy of Neutron Stars}
\label{sec:BE}

We calculate the core EOS assuming a minimal model of neutrons,
protons, electrons and muons in chemical equilibrium. We employ the
EOS of Baym, Pethick, and Sutherland~\cite{Baym:1971pw} for the
outer crust. Because the quantities we will be calculating are not
sensitive to the detailed composition of the inner crust, we simply
use polytropic EOSs that connect the uniform liquid core to the
bottom layers of the outer crust defined by the neutron dripline.
The stellar profiles are then determined by integrating the usual
Oppenheimer-Volkoff (OV) equations~\cite{Opp39_PR55}
\begin{eqnarray}
\frac{dP(r)}{dr}&=&-\frac{G}{c^{2}r^{2}}
  \bigg[\mathcal{E}(r)+P(r)\bigg]
  \left[M(r)+4\pi r^{3}\frac{P(r)}{c^{2}}\right]
  \left[1-\frac{2GM(r)}{c^{2}r}\right]^{-1},\\
\frac{dM_{\rm G}(r)}{dr}&=&4\pi r^{2}\frac{\mathcal{E}(r)}{c^{2}}
\label{Eq:OV}
\end{eqnarray}
where G is the gravitational constant, $P(r)$, $\mathcal{E}(r)$ and
$M(r)$ are the pressure, energy density and mass profiles as a
function of radial distance $r$. Given the boundary condition $P(0)
= P_{\rm c}$, $M(0)=0$, supplemented by an EOS, the radius of the
neutron star $R$ will then be determined from the condition of $P(R)
= 0$. The corresponding total mass enclosed is referred to as {\sl
the gravitational mass} of the neutron star, which is
\begin{eqnarray}
M_{\rm G}=\frac{1}{c^2}\int_{0}^{R} 4\pi r^{2}\mathcal{E}(r) dr,
\end{eqnarray}
One may also define the so-called {\sl baryonic mass} of the neutron
star as the mass of the star that has been disassembled into its
component baryons, which is given by $M_{\rm B}\equiv Nm_{\rm B}$.
Here $m_{\rm B}$ is the {\sl baryon mass} and $N$ is the total
number of baryons in the star that can be found by a volume
integration of the baryon density $\rho_{\rm B}(r)$ in the
Schwarzschild geometry,
\begin{eqnarray}
N=\int_{0}^{R}4\pi r^{2} \rho_{\rm B}
(r)\left[1-\frac{2GM(r)}{c^{2}r}\right]^{-1/2}dr,
\end{eqnarray}
Then the gravitational binding energy in mass units is simply
defined as $\mathcal{B}\equiv M_{\rm G}-M_{\rm
B}$~\cite{Weinberg:1972}, which is always a negative quantity for a
gravitationally bound system. It is also important to specify the
baryon mass $m_{\rm B}$. If we set it to be the mass of a single
nucleon, then $\mathcal{B}$ would be the total binding energy that
contains the contribution from the nuclear binding energy as well.
In order to compare the predicted binding energy with estimates of
the baryon mass of the pre-collapse cores, the nuclear binding
energy must be taken into account. It is therefore appropriate to
set the baryon mass to the atomic mass unit instead. Therefore
following
Refs.~\cite{Arnett:1976dh,Lattimer:2000nx,Podsiadlowski:2005ig,Newton:2009vz}
we set it to $m_{\rm B}\equiv931.5$ MeV$/c^{2}$. It his useful to
express the fractional gravitational binding energy
$\mathcal{B}/M_{\rm G}$ as a function of the compactness parameter
$\eta_{R}$~\cite{Lattimer:2000kb, Alecian:2003ez, Newton:2009vz},
which for neutron stars is several orders of magnitude larger then
that of regular stars such as our Sun~\cite{Lattimer:2000kb,
Eksi:2014wia}. As a result, the binding energy contribution to the
total mass of a neutron star is significant---up to $25\%$ of its
rest mass~\cite{Cook:1993qr, Alecian:2003ez, Lattimer:2010uk} for
stars near the maximum mass---although the exact size of the binding
energy strongly depends on the underlying EOS~\cite{Newton:2009vz}.

\begin{figure}[h]
\smallskip
\includegraphics[width=5in,angle=0]{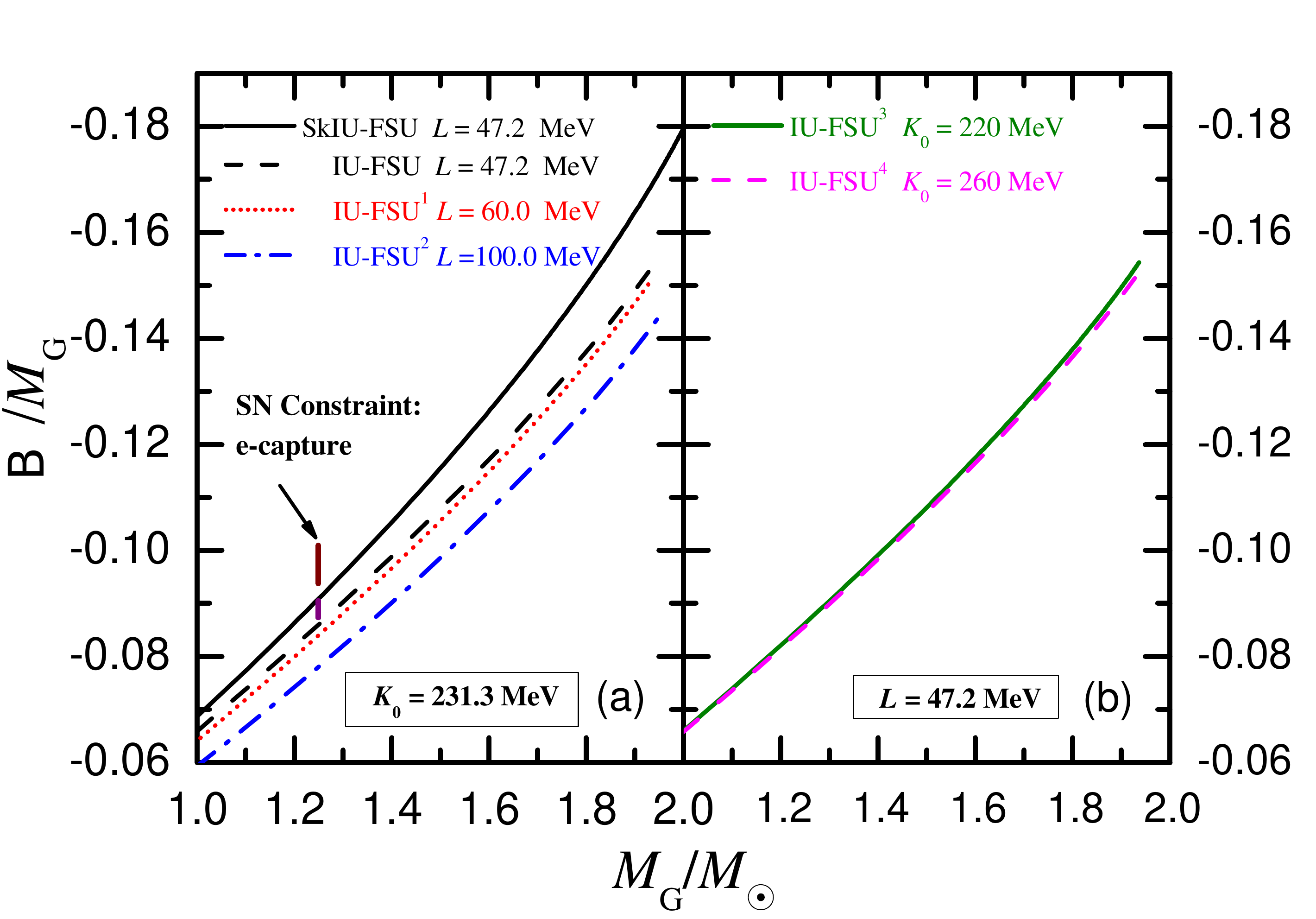}
\caption{(color online). The fractional gravitational binding energy
versus the gravitational mass of a neutron star for a set of the
EOSs discussed in the text: in the left panel (a) predictions are
given for the EOSs discussed in Fig.~\ref{Fig:EOS}, while in the
right panel (b) predictions are given for the IU-FSU-like models
with incompressibility coefficients of $K_0 = 220$ MeV (IU-FSU$^3$)
displayed in green solid line, and $K_0 = 260$ MeV (IU-FSU$^4$)
displayed in dashed magenta line. The constraints on the
gravitational binding energy of PSR J0737-3039B, under the
assumption that it is formed in an electron-capture supernova, are
given by the vertical brown and purple lines.} \label{Fig:BE}
\end{figure}

In Fig.~\ref{Fig:BE} we display the fractional gravitational binding
energy as a function of the total gravitational mass of a neutron
star. Higher mass stars, and stars governed by softer EOSs (that is,
softer symmetry energies or smaller values of incompressibility
$K_0$) are more compact and have higher absolute values of
$\mathcal{B}/M_{\rm G}$ (see also
\cite{Bagchi:2011sb,Podsiadlowski:2005ig}). Now let us examine
effects of the uncertain symmetry energy in more details. First,
notice that with almost the same low-density symmetry energy up to
about $1.5\rho_0$, but different high-density behaviors, predictions
for the $\mathcal{B}/M_{\rm G}$ in the SkIU-FSU and IU-FSU models
are very different. This effect becomes more pronounced with the
increase of the total gravitational mass. Indeed, larger neutron
stars probe the symmetry energy at higher densities which is
different in these two models due to their distinctly different
high-density behaviors of the symmetry energy. Recall that by
construction both the magnitude and the slope of the symmetry energy
at saturation are identical in these models, {\sl i.e.} $J = 31.3$
MeV and $L=47.2$ MeV, respectively. Choosing the IU-FSU  as a
reference, then the relative changes in the fractional gravitational
binding energies are $5.32\%$, $6.52\%$ and $9.89\%$ for
$1.25M_{\odot}$, $1.4M_{\odot}$ and $1.9M_{\odot}$ neutron stars
respectively (See Table.~\ref{Table:Sensitivity}). Thus, the
fractional gravitational binding energy becomes increasingly
sensitive to the the super-saturation behavior of the symmetry
energy for increasing mass.

This is in contrast to the behavior of the binding energy in models
with soft and stiff symmetry energies \emph{at saturation}---{\sl
e.g.}, the IU-FSU with $L=47.2$ MeV and the IU-FSU$^2$ with $L=100$
MeV. Again choosing the former as a reference model, then the
relative changes in the fractional gravitational binding energies
are $-10.26\%$, $-8.85\%$ and $-7.44\%$ for $1.25M_{\odot}$,
$1.4M_{\odot}$ and $1.9M_{\odot}$ neutron stars respectively. Thus
the fractional binding energy is most sensitive to the saturation
density slope of the symmetry energy for low mass neutron stars.

Finally, we note that, because the incompressibility  coefficient of
SNM, $K_{0}$, is presently well constrained by terrestrial nuclear
physics data, the effect on the fractional binding energy
$\mathcal{B}/M_{\rm G}$ over the constrained range of $K_{0}$ is
almost negligible.

The discovery of the double pulsar system PSR J0737-3039 has enabled
the accurate determination of many pulsar properties including both
pulsar's gravitational masses. In particular, the gravitational mass
of the pulsar PSR J0737-3039B is determined to be $M_{\rm G} =
1.2489 \pm 0.0007M_{\odot}$~\cite{Kramer:2006nb}, which is one of
the lowest reliably measured mass for any neutron star up to date.
Its low mass has led to the suggestion that this NS might have been
formed as a result of the collapse of an electron-capture supernova
from a progenitor of a O-Ne-Mg white
dwarf~\cite{Podsiadlowski:2005ig}. Under this assumption, it was
estimated that the baryonic mass of the precollapse O-Ne-Mg core
should lie between $1.366 M_{\odot} < M_{\rm B} <
1.375M_{\odot}$~\cite{Podsiadlowski:2005ig} and $1.358 M_{\odot} <
M_{\rm B} < 1.362M_{\odot}$~\cite{Kitaura:2005bt}. Using these two
sets of constraints, it was shown in Ref.~\cite{Newton:2009vz} that
the upper limit of the density slope at saturation should be $L
\approx 70$ MeV to be consistent with these observations. Notice
that recently an even lower mass of $M_{\rm G} = 1.230 \pm
0.007M_{\odot}$ has been reported through the long-term precision
timing measurements of double neutron star system PSR
J1756-2251~\cite{Ferdman:2014}. In Fig.~\ref{Fig:BE} we confirm that
only models with the soft symmetry energy are consistent with these
set of constraints. Notably, models with even smaller values of
$L$---such as IU-FSU$^1$ with $L=60$ MeV---could still be
inconsistent, if the subsequent density dependence of the symmetry
energy at higher densities is stiff. Such low mass stars, via their
binding energy, probe the symmetry energy slope at saturation; and,
once given a constrained range for $L$, they may also probe the
symmetry energy at higher densities.

Finally, we remark that the
fractional gravitational binding energy in the massive neutron star
PSR J0348+0432 is twice larger then that of the PSR J0737-3039B
independent of the EOS model used~\cite{Antoniadis:2013pzd}. As noted in that paper,
and shown in Fig.~\ref{Fig:BE}, whereas uncertainties of the EOS due to
variations of the density dependence of the symmetry energy are of
the order of $20\%$ for a given star, the substantial differences in
the $\mathcal{B}/M_{\rm G}$ of canonical and two-solar mass neutron
stars makes the latter an ideal laboratory for testing GR in the
strong field regime.

Thus we can conclude that, given that the EOS of SNM is well
constrained by terrestrial data, the binding energy of low mass
neutron stars probes the saturation density symmetry energy
primarily, and then the high density behavior of the symmetry
energy. Given constraints from low mass neutron stars, the resulting
range for the symmetry energy as a function of density constrains
$\mathcal{B}/M_{\rm G}$. For example, for the set of the EOSs used
here we find that the fractional binding energy of a 1.9 solar-mass
neutron star is between $-0.1380 \leq \mathcal{B} /M_{\rm G}\leq
-0.1639$. There is no known mechanism of extracting baryonic mass in
massive neutron stars however. In the next section we will analyze
this further by contrasting the effect of the density dependence of
the symmetry energy with measures of the strength of the
gravitational field.

\begin{widetext}
\begin{table}[h]
\caption{Predictions for the radius $R$, surface compactness
parameter $\eta_{R}$, fractional gravitational energy
$\frac{\mathcal{B}}{M_{\rm G}}$, full contraction of the Riemann
curvature tensor at the surface $\mathcal{K}_{R}$ of neutron stars
with masses 1.25$M_{\odot}$, 1.4$M_{\odot}$ and 1.9$M_{\odot}$ as
predicted by the various models used in the text. Also relative
changes of the gravitational binding energy and surface full
contraction of the Riemann curvature tensor with respect to the
predictions of the original IU-FSU model are provided in percentage.
Finally we give the radius, binding energy and curvature of a star
calculated using the scalar-tensor model described in the text is
given for the IU-FSU EOS (indicated by IU-FSU (ST)), together with
the relative change in binding energy compared with the general
relativistic result.} \label{Table:Sensitivity}
\begin{tabular}{l|c|c|c|c|c|c|c|c|c}
  \Xhline{2\arrayrulewidth}
  Model  &  $L$(MeV) & $K_0$(MeV) &$M_{\rm G}$($M_{\odot}$) & $R$(km) & $\eta_{R}$
  & $\frac{\mathcal{B}}{M_{\rm G}}$  & $\mathcal{K}_{R}$($10^{-2}\,{\rm km}^{-2}$)
  & $\frac{\Delta \mathcal{B}}{\mathcal{B}}(\%)$  & $\frac{\Delta \mathcal{K}_{R}}{\mathcal{K}_{R}}(\%)$\\
 \Xhline{2\arrayrulewidth}
             &   &  &  $1.25$  & 12.50 & 0.2953 & $-$0.0861  & 0.6544 & ---  & --- \\[-0.05cm]
 IU-FSU& 47.2   & 231.3 & $1.4$   & 12.49 & 0.3310 & $-$0.0989  & 0.7347 & ---  & --- \\[-0.05cm]
            &    &  & $1.9$  & 11.73 & 0.4785 & $-$0.1491  & 1.2051 & ---  & --- \\
  \hline
           &  &  & $1.25$   & 13.25 & 0.2786 & $-$0.0840  & 0.5492 & $-$2.48 & $-$19.15 \\[-0.05cm]
IU-FSU$^1$  & 60.0   & 231.3 & $1.4$  & 13.12 & 0.3152 & $-$0.0967  & 0.6345 & $-$2.21 & $-$13.64 \\[-0.05cm]
           &   &  & $1.9$  & 12.00 & 0.4676 & $-$0.1468  & 1.1242 & $-$1.57 & $-$~6.71 \\
  \hline
           &  &  & $1.25$   & 13.96 & 0.2644 & $-$0.0781  & 0.4697 & $-$10.26 & $-$39.32 \\[-0.05cm]
IU-FSU$^2$    & 100.0   & 231.3 & $1.4$   & 13.74 & 0.3010 & $-$0.0902  & 0.5524 & $-$8.85 & $-$24.81 \\[-0.05cm]
           &  & & $1.9$   & 12.44 & 0.4512 & $-$0.1380  & 1.0100 & $-$7.44 & $-$16.17 \\
  \hline
           &  & & $1.25$  & 12.50 & 0.2954 & $-$0.0863  & 0.6547 & $+$0.17 & $+$0.05 \\[-0.05cm]
IU-FSU$^3$           & 47.2   & 220.0 & $1.4$   & 12.48 & 0.3313 & $-$0.0991  & 0.7362 & $+$0.21 & $+$0.22 \\[-0.05cm]
           & & & $1.9$   & 11.70 & 0.4799 & $-$0.1496  & 1.2156 & $+$0.34 & $+$0.87 \\
  \hline
           &   &  & $1.25$   & 12.51 & 0.2951 & $-$0.0857  & 0.6533 & $-$0.44 & $-$0.18 \\[-0.05cm]
IU-FSU$^4$  & 47.2   & 260.0 & $1.4$   & 12.52 & 0.3304 & $-$0.0984  & 0.7307 & $-$0.51 & $-$0.54 \\[-0.05cm]
           &  &   & $1.9$   & 11.80 & 0.4757 & $-$0.1480  & 1.1838 & $-$0.77 & $-$1.77 \\
  \hline
           &   &  & $1.25$  & 11.81 & 0.3125 & $-$0.0909  & 0.7756 & $+$5.32 & $+$15.63 \\[-0.05cm]
SkIU-FSU  & 47.2   & 231.2 & $1.4$   & 11.71 & 0.3531 & $-$0.1054  & 0.8917 & $+$6.52 & $+$21.38 \\[-0.05cm]
           &  &  & $1.9$   & 10.85 & 0.5173 & $-$0.1639  & 1.5226 & $+$9.89 & $+$26.35 \\
     \hline
      IU-FSU (ST)   & 47.2   & 231.3 & $1.9$   & 12.29 & 0.4565 & $-$0.1509  & --- & $+$1.21  & --- \\
 \Xhline{2\arrayrulewidth}
\end{tabular}
\end{table}
\end{widetext}

\section{Curvature as the Strength of the Gravitational Field}
\label{sec:Curv}

In the absence of the cosmological constant the vacuum solution of
the Einstein's equations of GR, also known as the Schwarzchild
exterior solution, suggests that it is natural to measure the
strength of the gravitational field in terms of the compactness
parameter, $\eta \equiv 2GM/c^2r$. It can take any value between $0
< \eta < 1$, with $\eta = 0$ corresponding to the flat Minkowski
space in special relativity, while $\eta = 1$ corresponds to the
strongest gravitational field that can be observed by an observer at
infinity, the region which is known as the black hole event horizon.
However, as it was shown in Ref.~\cite{Psaltis:2008bb} the
compactness parameter $\eta$, which is directly related to the
Newtonian gravitational potential, could not serve as a fundamental
parameter in describing the gravitational field in GR. Indeed the
strength of the gravitational field is not characterized by a strong
field potential, but by a large curvature~\cite{Psaltis:2008bb}.
There are several quantities that can be used as a measure of the
curvature of gravitational field. These could, for example, be the
Ricci scalar, the Ricci tensor, and the Riemann tensor. However,
both the Ricci scalar and all components of the Ricci tensor vanish
in the exterior region of compact objects, which makes them
unsuitable choice for the measure of curvature. On the other hand,
there is a non-vanishing component of the Riemann tensor even in the
vacuum, ${\mathcal{R}^1}_{010} = - 2GM/c^2r^3 \equiv -\xi$, which
indicates that it should be a more relevant measure of curvature in
neutron stars~\cite{Psaltis:2008bb,Eksi:2014wia}. Since there are
totally 20 independent components of the Riemann tensor, it is best
that one uses the full contraction of the Riemann tensor instead,
which is a single scalar invariant. Therefore following the
Ref.~\cite{Eksi:2014wia} we use the square root of the Kretschmann
invariant, $\mathcal{K} \equiv
\sqrt{\mathcal{R}^{\mu\nu\rho\sigma}\mathcal{R}_{\mu\nu\rho\sigma}}$,
which is a full contraction of the Riemann tensor. For a spherically
symmetric mass configuration it can be expressed in terms of the
pressure $P(r)$, energy density $\mathcal{E}(r)$, and mass $M(r)$
profiles as~\cite{Eksi:2014wia},
\begin{equation}
\mathcal{K}^{2} =
\kappa^{2}\bigg[3\bigg(\mathcal{E}(r)+P(r)\bigg)^2-4\mathcal{E}(r)P(r)\bigg]
-\kappa\mathcal{E}(r)\frac{16GM(r)}{c^2r^3}+\frac{48G^{2}M^{2}(r)}{c^4r^6},
\label{Eq:Riemann}
\end{equation}
where $\kappa \equiv 8\pi G/c^{4}$. One can then readily evaluate
the Kretschmann invariant while integrating the OV equations. Notice
that outside the star this curvature has a desired form,
$\mathcal{K} \equiv 2\sqrt{3} \xi$.

\begin{figure}[h]
\smallskip
\includegraphics[width=5in,angle=0]{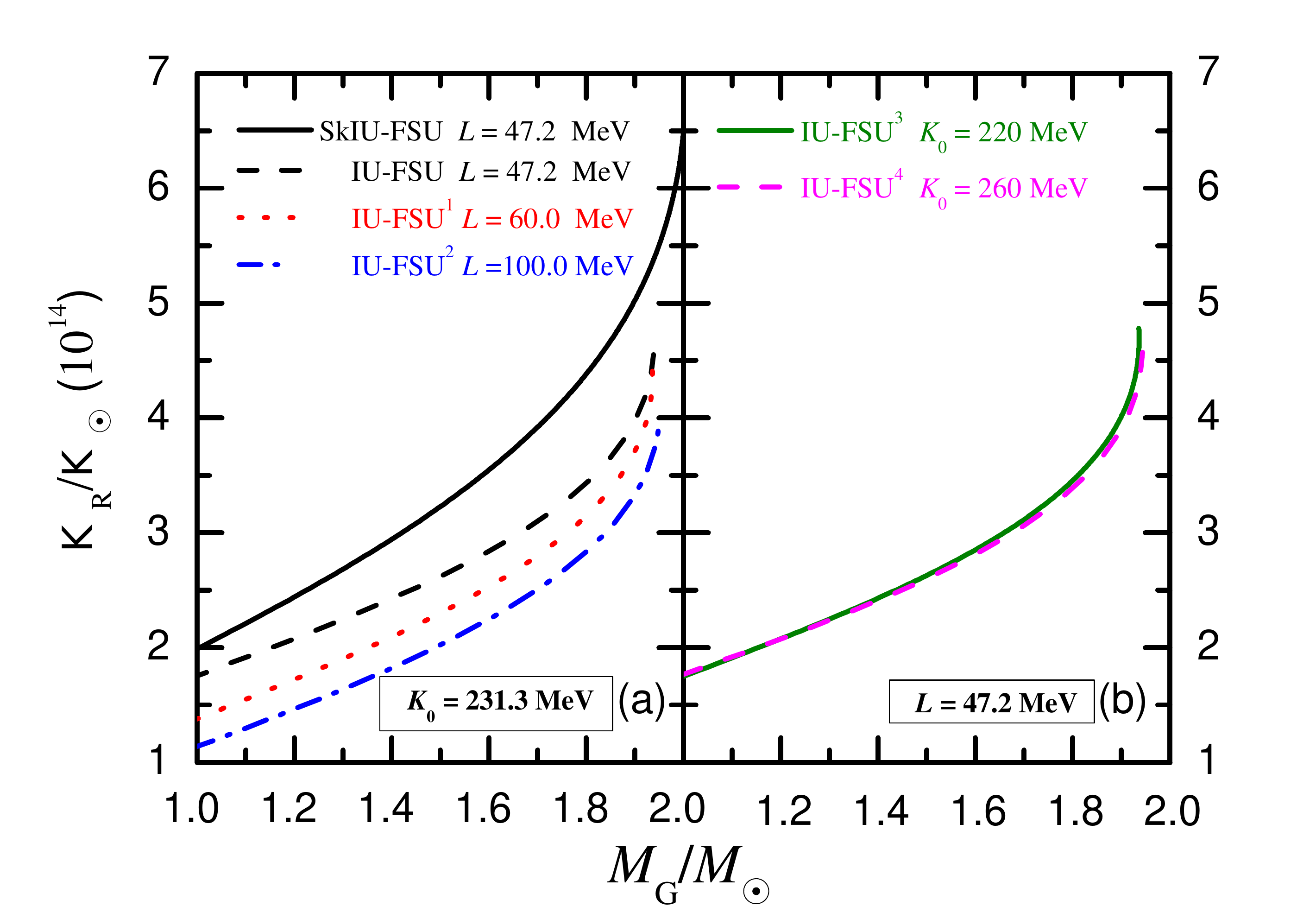}
\caption{(color online). The ratio of surface curvatures of neutron
stars and the Sun, $\mathcal{K}_{R}/\mathcal{K}_{\odot}$, as a
function of the neutron stars gravitational mass for the EOSs
considered in this work. Left panel (a) represents the uncertainty
range of the nuclear symmetry energy, while the right panel (b)
represents the uncertainty range of the incompressibility $K_0$ of
SNM.} \label{Fig:KR}
\end{figure}

It is common to measure the strength of gravity in neutron stars
with respect to that of the solar system. At the surface of the Sun,
the curvature is equal to $\mathcal{K}_{\odot} \equiv 3.0 \times
10^{-17}$ km$^{-2}$, which may be regarded as a {\sl small}
quantity. On the other hand, neutron stars can have surface
curvature of as large as 14 orders of magnitude greater. In
Fig.~\ref{Fig:KR} we display the ratio of the neutron-star surface
curvature to the one of the Sun,
$\mathcal{K}_{R}/\mathcal{K}_{\odot}$, as a function of the neutron
star mass. Clearly, this ratio becomes even larger for more massive
neutron stars. Notice that for less massive stars the uncertainty in
the nuclear symmetry energy has a large effect on the curvature;
with variation of  $\sim 50\%$ in the surface curvature among EOSs
for stars less than $1.4M_{\odot}$. Depending on the symmetry
energy, a surface curvature of $\mathcal{K}_{R} = 2.0 \times 10^{14}
\, \mathcal{K}_{\odot}$ can be obtained for neutron stars with
masses in the range of $1.0 < M_{\rm G}/M_{\odot} < 1.5$. The
knowledge about the symmetry energy thus becomes crucial in testing
various models of strong-field gravity. The effects of the symmetry
energy are less pronounced on the curvature of massive neutron
stars, but still significant. A curvature of $\mathcal{K}_{R} = 4.0
\times 10^{14} \, \mathcal{K}_{\odot}$ have been obtained for
massive neutron stars in the range of $1.72 < M_{\rm G}/M_{\odot} <
1.95$. Quantitatively, the relative variation in $\mathcal{K}_{R}$
of a canonical 1.4 solar-mass NS predicted by the IU-FSU$^{2}$ and
SkIU-FSU is $61\%$, and this change reduces to only $51\%$ for a 1.9
solar-mass NS. This is comparable to the relative change between NSs
with masses $1.4M_{\odot}$ and $1.9M_{\odot}$ within the two models:
$83\%$ and $71\%$ for the IU-FSU$^{2}$ and SkIU-FSU, respectively
(See Table~\ref{Table:Sensitivity}).

Coupled with the sudden increase in the surface curvature of neutron
stars close to their maximum mass, this emphasizes the fact that
lower mass neutron stars are more natural laboratories for testing
the EOS, and that only once the EOS is significantly constrained can
high mass neutron stars come into their own as laboratories for
testing alternative models for gravity.

In the same spirit as in the previous section, we now compare the
effect of the uncertainties due to the variations of density
dependence of the symmetry energy at saturation density and at high
densities. There is a $25\%$ reduction in the value of the surface
curvature of a $1.4 M_{\odot}$ NS, when we compare models with the
soft and stiff symmetry energies as controlled  by their density
slope at saturation of $L = 47.2$ MeV (IU-FSU) and $L=100$ MeV
(IU-FSU$^2$), correspondingly (See Table~\ref{Table:Sensitivity}).
Obviously, the curvature is much larger when the soft symmetry
energy is used.

This can be compared with a reduction in stiffness in the
high-density EOS: comparing predictions of the SkIU-FSU and IU-FSU,
we see that the surface curvature further increases by about $21\%$
for a $1.4 M_{\odot}$ NS and $26\%$ for a $1.9 M_{\odot}$ NS. Thus
the surface curvature is significantly more sensitive than the
gravitational binding energy to the variation of the density
dependence of the symmetry energy at both saturation densities and
higher. Again, its variation with the incompressibility of SNM is
negligible (See Table~\ref{Table:Sensitivity} and
Fig.~\ref{Fig:KR}).

\begin{figure}[h]
\smallskip
\includegraphics[width=5in,angle=0]{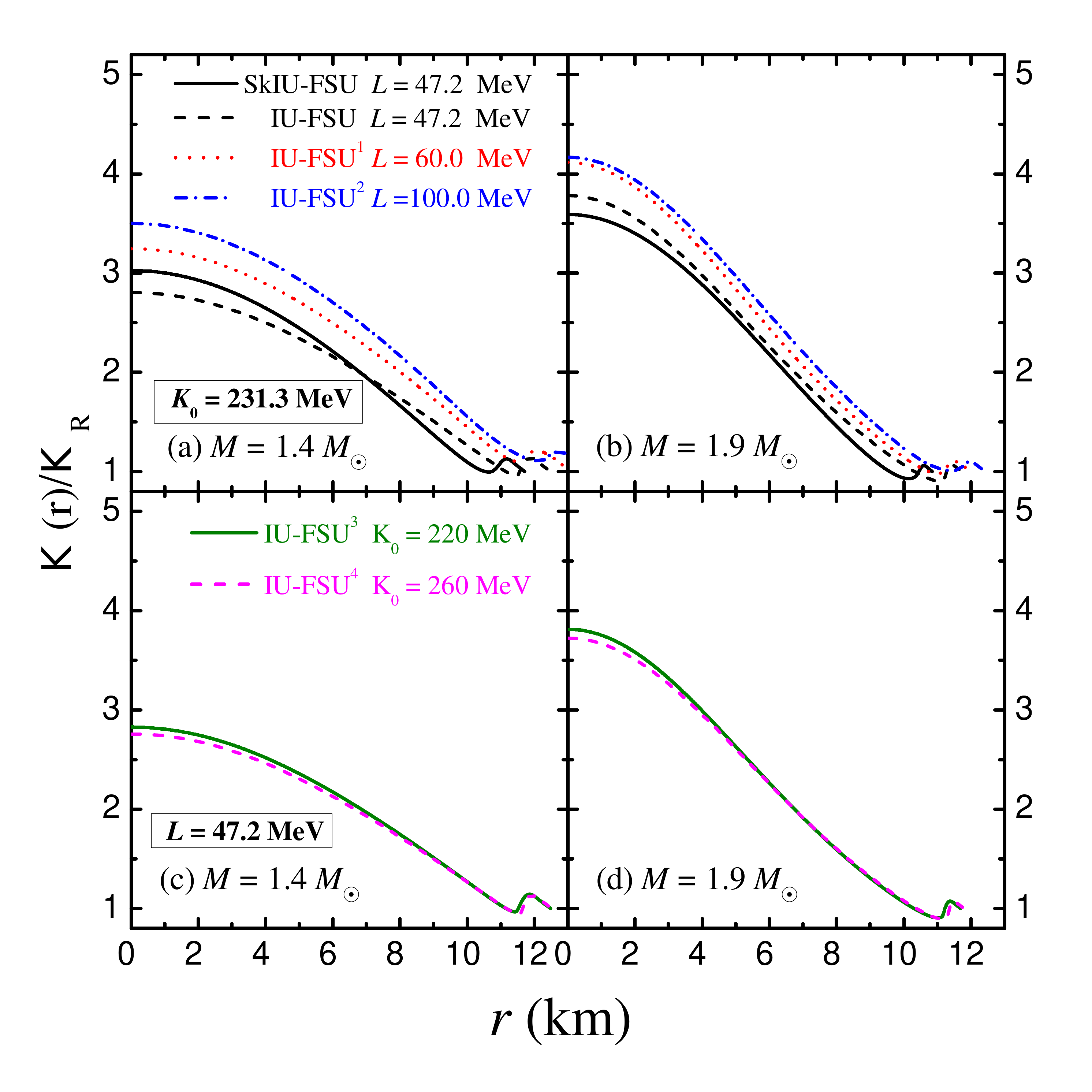}
\caption{(color online). Normalized curvature profiles
$\mathcal{K}(r)$ with respect to its value at the surface of the
star $\mathcal{K}_{R}$ for the EOSs considered.}
\label{Fig:K}
\end{figure}

We next consider the strength of the gravitational field within the
neutron star. Notice that towards the center of the star the
compactness $\eta$ would approach zero. On the contrary, the
curvature stays non-zero throughout the neutron
star~\cite{Eksi:2014wia}. In order to study the strength of the
gravitational field within the NS, we normalize the curvature
profile $\mathcal{K}(r)$ with respect to its value at the surface of
the star $\mathcal{K}_{R}$. The normalized profile tells us how many
times stronger the curvature $\mathcal{K}(r)$ gets as one approaches
the central region. This also provides a convenient way to compare
effects of the density dependence of the symmetry energy on the same
basis. In Fig.~\ref{Fig:K} we plot the curvature profiles for a 1.4
and 1.9 solar-mass neutron stars using the EOSs considered in this
work. For a canonical $1.4M_{\odot}$ NS the central curvature is
$2.8$ to $3.5$ times stronger then the surface curvature depending
on the EOS used. This ratio becomes larger for massive neutron
stars, and in particular for a 1.9 solar mass NS it can be as large
as $3.6$ to $4.2$ times stronger. Notice that the ratio becomes
especially larger for models with the stiff saturation-density
symmetry energy (See Fig.~\ref{Fig:K}), whereas the high density
behavior of the symmetry energy has a significantly smaller effect
on the ratio of curvatures. Again, we observe a very small effect
due to the variation of $K_{0}$.

Fig.~\ref{Fig:KR} and Fig.~\ref{Fig:K} together give us a complete
picture of the curvature $\mathcal{K}(r)$ throughout the neutron
star. First of all, the curvature of massive neutron stars at the
surface is much larger then that of canonical NSs. Secondly, as one
moves towards the center the strength of gravity becomes even
stronger. Whereas this explicitly demonstrates the important role of
massive neutron stars in studying the strong-field gravity, the
uncertainties of the EOS resulting from our poor knowledge of the
density dependence of the symmetry energy cannot be overlooked. It
was put forward in Ref.~\cite{Antoniadis:2013pzd} that the two solar
mass neutron star in PSR J0348+0432---which is the most massive
neutron star observed up to date---can be used as a testbed for
gravity in the strong field regime. This conclusion was based on the
fact that its binding energy is far outside the presently tested
binding energy range. While supporting this conclusion, we have also
analyzed the effects of the density dependence of the symmetry
energy on the strength of the gravitational field and have found
that current uncertainties in the EOS are still very large to
cleanly probe the fundamental physics in the extreme conditions.
Nevertheless, massive NSs that are manifest as radio pulsars remain
valuable probes of both models of gravity and theories of dense
matter in the extreme conditions that are inaccessible to the
terrestrial experiments. In the next section we will analyze our
results for the case of a neutron star in the scalar-tensor
theories.

\section{Neutron Stars in a Scalar-Tensor Model of Gravity}
\label{sec:ScalarTensor}

Motivated by the fact that the properties of neutron stars are not
only sensitive to the underlying EOS, but also to the strong-field
behavior of gravity, in this section we will consider neutron stars
in the simplest natural extension of the GR known as the {\sl
scalar-tensor theories} of gravity. According to these theories, in
addition to the second-rank metric tensor, $g_{\mu\nu}$, the
gravitational force is also {\sl mediated} by a scalar field,
$\varphi$. The most general form of the action defining this theory
is written as~\cite{Damour:1996ke,Yazadjiev:2014cza}
\begin{equation}
S = \frac{c^4}{16 \pi G} \int d^4x \sqrt{-g^{\ast}}\left[R^{\ast} -2
g^{\ast \mu \nu} \partial_{\mu} \varphi \partial_{\nu} \varphi -
V(\varphi)\right] + S_{\rm matter}\left(\psi_{\rm matter};
A^2(\varphi)g^{\ast}_{\mu\nu} \right) \ ,
\end{equation}
where $R^{\ast}$ is the Ricci scalar curvature with respect to the
so-called {\sl Einstein frame} metric $g^{\ast}_{\mu\nu}$ and
$V(\varphi)$ is the scalar field potential. The physical metric,
also known as {\sl Jordan frame metric}, is defined as $g_{\mu\nu}
\equiv A^2(\varphi)g^{\ast}_{\mu\nu}$. Notice that the GR is
automatically recovered in the absence of the scalar field. The
relativistic equations for stellar structure then can be written as
\begin{eqnarray}
\frac{dM_{\rm G}(r)}{dr} &=& \frac{4 \pi}{c^2} r^2 \mathcal{E}(r)
A^4(\varphi) +
\frac{r^2}{2}\left(1-\frac{2GM(r)}{c^2r}\right)\chi^2(r) +
\frac{r^2}{4} V(\varphi) \ , \\
\frac{dM_{\rm B}(r)}{dr} &=& 4\pi m_{\rm B} r^2 \rho(r) A^3(\varphi) \ , \\
\frac{d \varphi(r)}{dr} &=& \chi(r) \ , \\
\nonumber \frac{d \chi(r)}{dr} &=& \Bigg[1-\frac{2 G M(r)}{c^2
r}\Bigg]^{-1}\Bigg\{\frac{4 \pi G}{c^4} A^4(\varphi)
\bigg[\alpha(\varphi) \bigg(\mathcal{E}(r) - 3 P(r)\bigg)+r\chi(r)\bigg(\mathcal{E}(r) - P(r)\bigg)\bigg] \ \\
&-& \frac{2}{r}\left(1-\frac{ G M(r)}{c^2 r}\right) \chi(r)
+\frac{1}{2}r\chi(r)V(\varphi) +
\frac{1}{4}\frac{dV(\varphi)}{d\varphi}\Bigg\} \ , \\
\nonumber \frac{dP(r)}{dr} &=& -\bigg(\mathcal{E}(r) +
P(r)\bigg)\Bigg[1-\frac{2 G M(r)}{c^2 r}\Bigg]^{-1}\Bigg\{ \frac{4
\pi G}{c^4} r A^4(\varphi) P(r)  + \frac{G}{c^2r^2}M(r)  \ \\
&+& \left(1-\frac{2 G M(r)}{c^2 r}\right)\left(\frac{1}{2}r\chi^2(r)
+ \alpha(\varphi)\chi(r)\right) -\frac{1}{4}rV(\varphi) \Bigg\} \ ,
\end{eqnarray}
where
\begin{equation}
\alpha(\varphi) \equiv \frac{\partial \ln A(\varphi)}{\partial
\varphi} \ .
\end{equation}
At the limiting case of $A(\varphi) = 1$, and $V(\phi) = 0$ one
recovers the general relativistic OV Eqs.~(\ref{Eq:OV}). Following
the Ref.~\cite{Damour:1996ke} we set $V(\varphi)=0$ which can only
appear in models of modified gravity, and consider a coupling
function of the form
\begin{equation}
A(\varphi) = \exp\left(\alpha_0\varphi + \frac{1}{2}\beta_0
\varphi^2\right) \ .
\end{equation}
Again once an EOS is supplemented, for a given central pressure
$P(0) = P_{\rm c}$ one can then integrate the equations above from
the center of the star to $r \rightarrow \infty$. At the center of
the star the Einstein frame boundary conditions are given as
\begin{equation}
P(0) = P_{\rm c} \ , \qquad \mathcal{E}(0) = \mathcal{E}(\rm c) \ ,
\qquad \varphi(0) = \varphi_{\rm c} \ , \qquad \chi(0) = 0 \ ,
\end{equation}
while at infinity we demand cosmologically flat solution to agree
with the observation
\begin{equation}
\lim_{r \rightarrow \infty} \varphi(r) = 0 \ .
\end{equation}
The stellar coordinate radius is then determined by the condition of
$P(r_{\rm s}) = 0$. The physical radius of a neutron star is found
in the Jordan frame as $R_{\rm NS} = A^2\left[\varphi(r_{\rm
s})\right]r_{\rm s}$. Notice however that the physical stellar mass
as measured by an observer at infinity, also known as the {\sl
Arnowitt-Deser-Misner} (ADM) mass $M_{\rm ADM}$, matches with the
coordinate mass, since at infinity the coupling function approaches
unity. It was shown in~\cite{DeDeo:2003ju} that measurement of the
surface atomic line redshifts from neutron stars could be used as a
direct test of strong-field gravity theories. In particular, as a
representative of the scalar-tensor theory a model with $\alpha_0=0$
and $\beta_0=-8$ was used. Notice that however a significant
improvement have been made since then in constraining the $\alpha_0$
and $\beta_0$ parameters. For example, very recently the authors of
Ref.~\cite{Freire:2012mg} put the most stringent constraint on these
parameters from the observations of the 10-year timing campaign on
PSR J1738+0333. According to these latest constraints the quadratic
parameter should take values of $\beta_0 \gtrsim -5.0$. Moreover, as
shown first in Ref.~\cite{Damour:1996ke} predictions by models with
$\beta_0>-4.35$ can not in general be distinguished from the general
relativistic results---{\sl i.e.} from the model with
$\{\alpha_0,\beta_0\}= \{0,0\}$---due to the so-called ``spontaneous
scalarization" effect. Since we have already showed that the
uncertainties in the SNM EOS due to the variation of $K_0$ has
insignificant effects, in this section we will solely concentrate on
the effects of the symmetry energy. Further, we will choose the
upper bounds on the parameters of the scalar-tensor theory to be
$\alpha_0^2 < 2.0 \times 10^{-5}$ and $\beta_0 > -5.0$ as
constrained by the observation~\cite{Freire:2012mg}.

\begin{figure}[h]
\smallskip
\includegraphics[width=5in,angle=0]{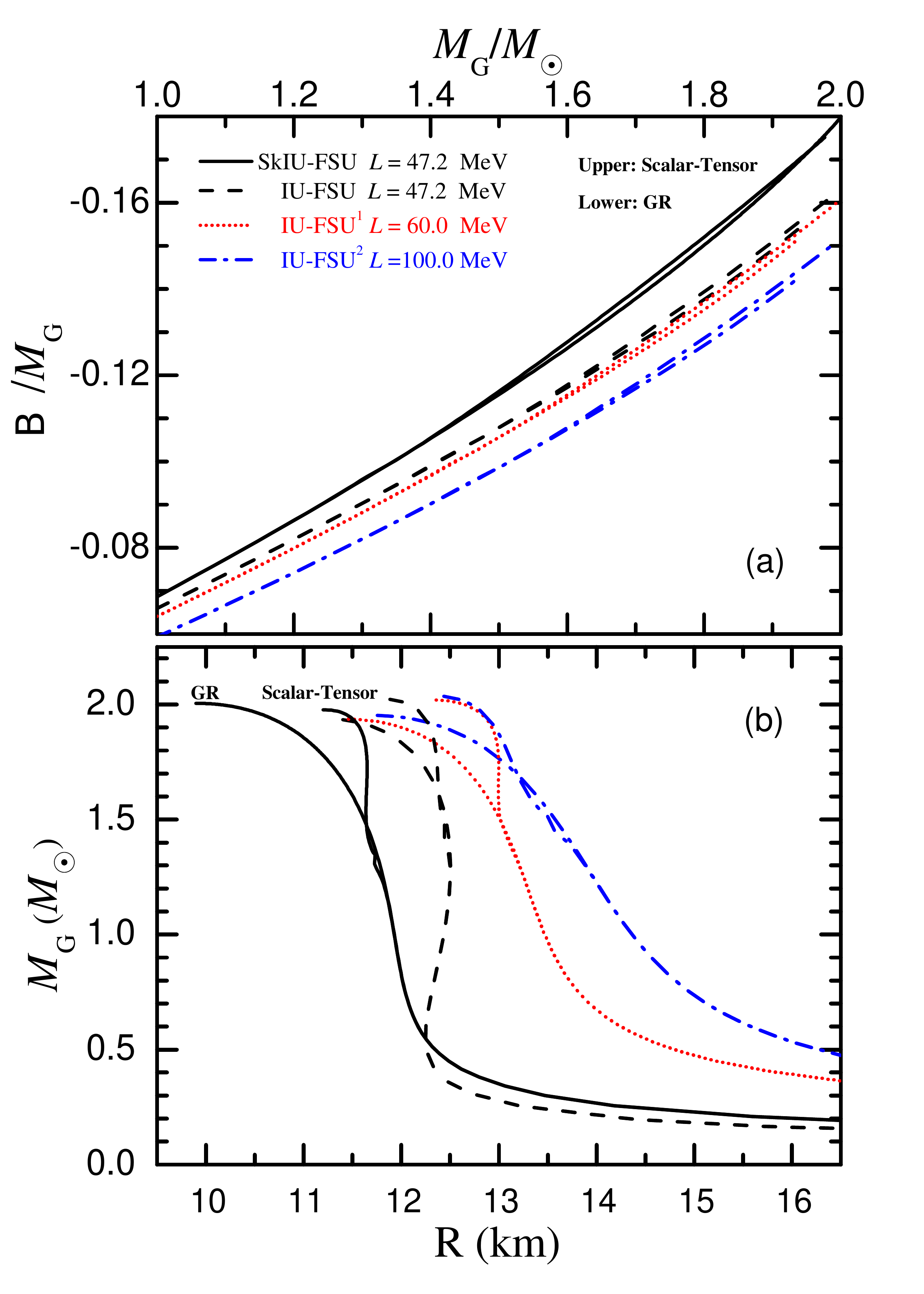}
\caption{(color online). The fractional gravitational binding energy
(a) and the mass versus radius relation (b) calculated using the
EOSs considered in this work. The upper observational bound on the
scalar-tensor theory parameters of $\{\alpha_0,\beta_0\} =
\{\sqrt{2.0 \times 10^{-5}}, -5.0\}$ have been used.} \label{Fig:ST}
\end{figure}

In Fig.~\ref{Fig:ST} we present our results for the fractional
gravitational binding energy as a function of the stellar mass
(Fig.~\ref{Fig:ST}a) and the mass versus radius relation
(Fig.~\ref{Fig:ST}b). The general relativistic (GR) predictions are
then compared with the predictions of the scalar-tensor theories
using different EOSs. We observe that the GR predictions give a
lower absolute value of the fractional binding energy for a given
stellar mass then the scalar-tensor theory, in general. However the
difference is quite negligible as far as the uncertainties in the
EOS is concerned. Moreover, low-mass neutron-stars are
indistinguishable in these two models of gravity, because the
critical value for the so-called ``spontaneous scalarizaton" occurs
only when  NSs exceed masses of about $1.4$ solar mass. To further
illuminate the effects of compactness on the binding energy we can
look at the mass versus radius relation. We observe that there is a
slight change in this relation; notably, massive neutron stars have
smaller radii in GR. If one uses a different value of $\beta_0=-8.0$
as in the Ref.~\cite{DeDeo:2003ju}---which {\sl has already been
excluded} by the observation---one can obtain very massive neutron
stars. In particular, the IU-FSU EOS gives a maximum mass of
$3.2M_{\odot}$ and a slightly larger radius of $14.5$ km. However,
under present observational constraints, differences in predictions
using the GR and the scalar-tensor theories are much smaller than
those due to the uncertainties in the EOS. Thus measurements of
gravitational binding energy and neutron star radii will lead to
significant constraints on the neutron star symmetry energy at
saturation and high densities, rather than constraining gravity
models between GR and scalar-tensor theories.

\section{Future determination of the nuclear symmetry energy from the surface gravitational redshift and the maximum spin frequency}
\label{sec:AstrophObserv}
\begin{figure}[h]
\smallskip
\includegraphics[width=5in,angle=0]{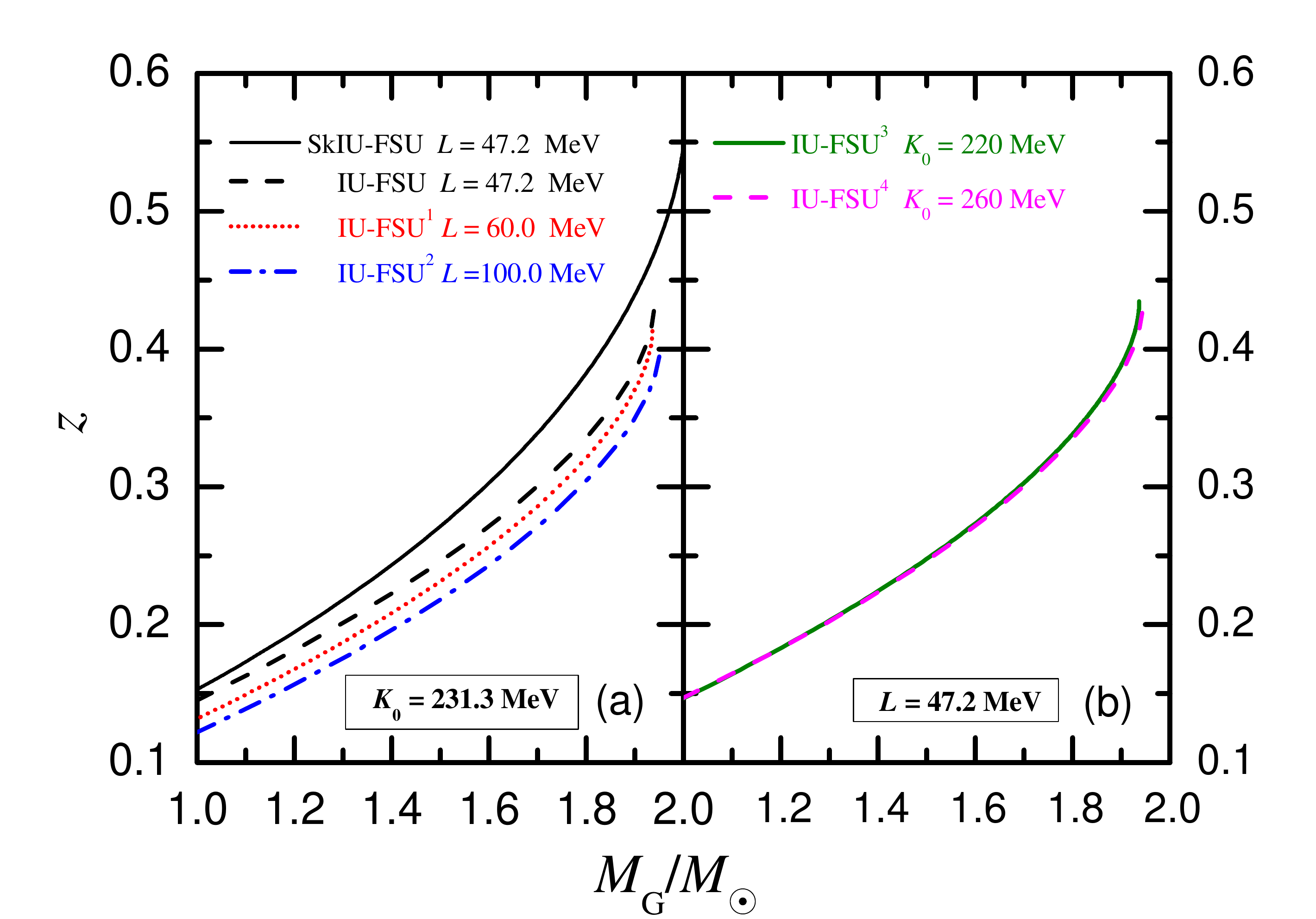}
\caption{(color online). The effect of the density dependence of the
nuclear symmetry energy (a) and the incompressibility coefficient
(b) on the surface gravitational redshift $z$ of neutron stars.}
\label{Fig:Redshift}
\end{figure}
Having established that uncertainties in the EOS arising from the nuclear
symmetry energy outweigh those within the set of Scalar-Tensor gravity theories,
we now seek in this section for
some possible astrophysical measurements (see
Ref.~\cite{Miller:2013tca} for review) that may help constrain the
density dependence of nuclear symmetry energy. In Fig.~\ref{Fig:BE} we showed the
binding energy, which can be inferred from the masses of binary pulsars assuming a specific evolutionary pathway
for which strong (albeit circumstantial) evidence exists for two systems in particular;
currently such inferences favor a softer symmetry energy \cite{Podsiadlowski:2005ig,Newton:2009vz}.
It was shown by some
of us previously that the pulsar moment of inertia is an observable sensitive
to the density dependence of the nuclear symmetry
energy~\cite{Worley:2008cb,Fattoyev:2010tb}. In particular, it was
shown that knowledge of the moment of inertia for pulsar PSR
J0737-3039A  with a 10\% accuracy could help discriminate among
various equations of state. Moreover, it was recently shown by some
of us that the observed symmetry energy effect on the tidal
polarizability (also known as tidal deformability) parameter in
binary neutron stars is very strong~\cite{Fattoyev:2012uu}, and in
particular it has been shown that future measurement of the tidal
polarizability in binary mergers would stringently constrain the
high-density behavior of the symmetry energy. Perhaps the simplest
of all, the knowledge of stellar mass and radius even from a single
neutron star would provide very useful information on the equation
of state of neutron star matter as first pointed out by
Ref.~\cite{Lindblom:1992}. While the mass of the neutron star can be
measured fairly accurately in binary systems, the measurement of its
radius is usually very complicated. From the observational point of
view, it is more relevant to describe stellar properties with the
so-called {\sl radiation radius},
\begin{equation}
R_{\infty} =\frac{R}{\sqrt{1-\eta_R}} \ ,
\end{equation}
and the surface gravitational redshift,
\begin{equation}
z = \frac{1}{\sqrt{1-\eta_R}}  - 1 \ .
\end{equation}
Notice that the expression for redshift is only valid in the
slow-rotation approximation, where the Doppler effect due to the
rotation of the star and the Lense-Thirring effect have been
neglected. The radiation radius $R_{\infty}$ could be obtained from
a combination of flux and temperature measurements from the neutron
star's surface and distance to the source. There are however some
major uncertainties involved in the determination of $R_{\infty}$
such as the distance, interstellar absorption, and details
concerning the composition of the atmosphere and its magnetic field
strength and structure \cite{Lattimer:2006xb}. Another possible
combination to uniquely determine the stellar mass and radius could
come from the inference of the surface gravitational acceleration
and the gravitational redshift as pointed out by
Ref.~\cite{Majczyna:2006tv}. The gravitational redshift can be
measured from the possible shift of atomic absorption lines in
spectra of the stars. One of the main difficulties in measuring the
gravitational redshift of signals emitted from the surface of
neutron stars is due to the existence of strong surface magnetic
field. Indeed, a value of $z=0.35$ was reported~\cite{Cottam:2002cu}
based on an analysis of the stacked bursts in the low-mass X-ray
binary EXO 0748-676. Unfortunately, subsequent observations could
not confirm the existence of such spectral
lines~\cite{Cottam:2007cd}. Nevertheless, future developments of
large-area instruments in X-ray spectroscopy missions could make it
possible to detect the spectral lines from the
surface~\cite{Bhattacharyya:2004nb}.

\begin{figure}[h]
\smallskip
\includegraphics[width=5in,angle=0]{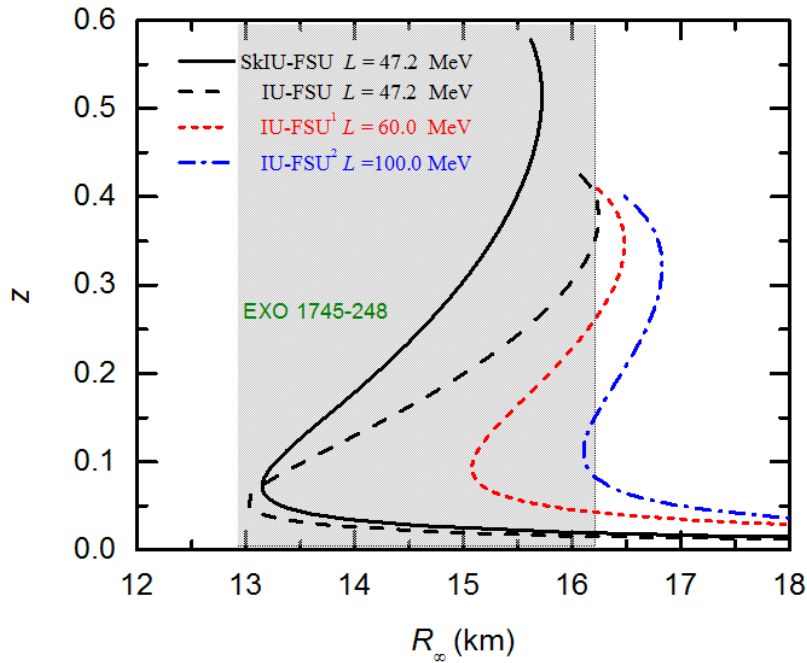}
\caption{(color online). Surface gravitational redshift $z$ as a
function of stellar radiation radii $R_{\infty}$ are constructed
from the four EOSs discussed in the text.} \label{Fig:Redshift2}
\end{figure}

In Fig.~\ref{Fig:Redshift}, we plot the surface gravitational
redshift as a function of the stellar mass for various equations of
state discussed in previous sections. As it is evident from the
figure, the effect of the density dependence of the nuclear symmetry
energy is quite significant. For a canonical neutron star,
uncertainties resulting from the density dependence of the symmetry
energy would allow a range of predictions for the surface redshifts
spanning from $0.197 < z < 0.244$. Interestingly, had the $z=0.35$
redshift in EXO 0748-676 been confirmed, its mass would then have to
lie in the range of $1.71 M_{\odot}< M < 1.89 M_{\odot}$ similar to
that found earlier in \cite{Li:2005sr}. On the other hand, if future
redshift measurements report $z>0.6$ then all current models of the
equation of state would be ruled out. Obviously, the effect of
rotation would significantly alter the predictions for the redshift,
and therefore should be taken into account. Such a study is in
progress and results will be reported elsewhere.

Next, we plot in Fig.~\ref{Fig:Redshift2} the surface
gravitational redshift $z$ as a function of the radiation radius
$R_{\infty}$ by varying only the density dependence of the symmetry energy. The radiation
radius of the low-mass X-ray source, EXO 1745-248 has been
determined to be $R_{\infty} = 14.57 \pm 1.64$ km (shaded band) ~\cite{Ozel:2008kb, Lattimer:2014sga}.
It is seen that the IU-FSU model with $L =100$ MeV is
almost excluded by this observation regardless of the gravitational
redshift for this source. Future precise determination of the
radiation radius from the low-mass X-ray binaries would therefore
more stringently constrain the density dependence of the symmetry
energy.

\begin{figure}[h]
\smallskip
\includegraphics[width=5in,angle=0]{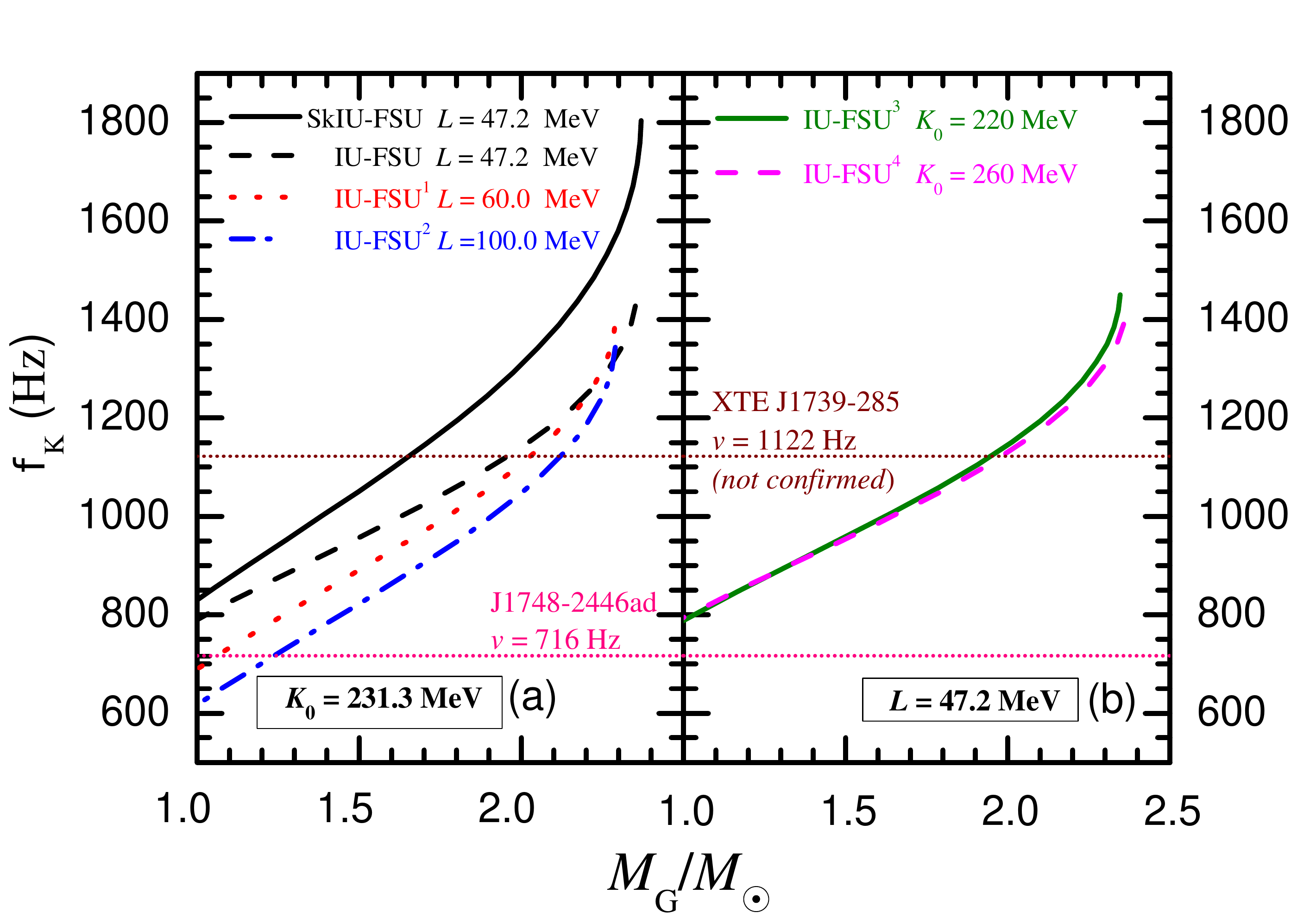}
\caption{(color online). Maximum spin frequencies as a function of
stellar masses obtained calculated for the four EOSs discussed in
the text using the publicly available code for rapidly rotating
neutron stars~\cite{Stergioulas_RNS}.} \label{Fig:MaxFrequency}
\end{figure}

A certain combination of the stellar mass and radii could
alternatively be constrained by measuring the spin frequencies. The
maximum spin frequency for a stellar structure is known as the
Kepler frequency, which in Newtonian gravity can be calculated using
$f_{\rm K} = \frac{1}{2 \pi}\sqrt{\frac{G M}{R^3}}$. The general
relativistic expression for the maximum spin frequency is very
complicated and can be found by solving an equation involving metric
functions and their derivatives~\cite{Weber:1999}. As first shown in
Ref.~\cite{Haensel:1989}, the maximum spin frequency for a maximum
allowable mass configuration can be analytically written in terms of
the stellar properties as $f_{\rm K, max} = C \left(\frac{M_{\rm
max}}{M_{\odot}}\right)^{1/2}\left(\frac{R_{M_{\rm max}}}{10 {\rm
km}}\right)^{-3/2}$, where the $M_{\rm max}$ and $R_{M_{\rm max}}$
are the maximum allowable mass and the corresponding radius for
static configurations. The most recent and updated analysis gives a
fitting coefficient of $C = 1.08$ kHz~\cite{Haensel:2009}. While
this expression is highly accurate in estimating the maximum spin
frequency, it is limited to the maximum mass configuration only.
Moreover, rotating stars assume an oblate shape thus allowing a
larger maximum mass than their static counterparts for a given
equation of state. Therefore, without relying on the approximate
expression we have used the {\verb"RNS"} code for rapidly rotating
neutron stars~\cite{Stergioulas_RNS}. In Fig.~\ref{Fig:MaxFrequency}
we show the maximum spin rate as a function of the stellar mass. The
mass of the fastest spinning pulsar known today J1748-2446ad---with
a frequency of $f = 716.356$ Hz---is currently
unknown~\cite{Hessels:2006ze}. Should it be a low-mass neutron star
with $M < 1.29 M_{\odot}$, models with larger density slope of the
symmetry energy at saturation would be excluded. While not confirmed
by subsequent observations, a weak evidence for an $f = 1122$ Hz
signal has been reported during thermonuclear bursts from the XTE
J1739–285 neutron star~\cite{Kaaret:2006gr}.  Although this
observation does not have strong statistical significance, our
calculations indicate that its mass should be above $1.65 M_{\odot}
< M < 2.12 M_{\odot}$ depending on the EOS model (see
Fig.~\ref{Fig:MaxFrequency}). Conversely, if such a fast-spinning
neutron star had been found to have a mass of $M = 1.7 M_{\odot}$ or
less, then only models with the very soft nuclear symmetry energy
such as the SkIU-FSU would have survived. Although at this stage
there is no known pulsars with larger spin frequencies, future
measurements of such frequencies would be decisive in ruling out
models predicting the stiff nuclear symmetry energy.

\section{Summary}
\label{sec:Conclusion} There is a degeneracy between the EOS for
super-dense matter and the strong-field gravity in understanding
properties of neutron stars. Our goal is to provide information that
may help break this degeneracy. Since the most uncertain part of the
nucleonic EOS is currently the density dependence of nuclear
symmetry energy especially at super-saturation densities, we have
studied effects of the nuclear symmetry energy within its current
uncertain range on the binding energy and curvature of neutron stars
within GR and the scalar-tensor theory of gravity. We found that the
gravitational binding energy is moderately sensitive to the
underlying symmetry energy, with lower mass neutron stars $\lesssim
1.4M_{\odot}$ probing primarily the saturation-density stiffness of
the symmetry energy, and higher mass stars being more sensitive to
the high density behavior of the symmetry energy. For the most
massive NSs discovered so far, we found an almost 20\% change in
their binding energies by varying the density dependence of the
symmetry energy within its uncertainty range determined by recent
terrestrial nuclear laboratory experiments. We also found that the
binding energy is quite insensitive to the variation of the
incompressibility $K_0$ of SNM EOS.

The curvature of neutron stars measured using the square root of the
full contraction of the Riemann tensor was found to be significantly
more sensitive to the variation of the density dependence of nuclear
symmetry energy but not to that of the incompressibility $K_0$, with
variations in surface curvature of order $50\%$ over the whole mass
range.

Within the scalar-tensor theory of gravity which is the simplest
natural extension of the GR,  using upper bounds on the parameters
of the scalar-tensor theory from observation, we find negligible
change in the binding energy of neutron stars over the whole mass
range, and significant changes in radii only for neutron stars well
above $1.4M_{\odot}$. Even then, the changes in radii are smaller
than those that result from variation of the stiffness of the
symmetry energy.

Restricting ourselves to the scalar-tensor modifications to gravity,
we conclude that astrophysical measurements of quantities related to
the compactness and curvature of neutron stars such as the
gravitational binding energy, atomic line redshifts, and the maximum
spin frequency will help constrain the neutron star EOS further
rather than the model of gravity. However, the large range of
curvatures predicted by varying the symmetry energy may still mean
that current uncertainties in the density dependence of nuclear
symmetry energy are still too large to clearly break the EOS-gravity
degeneracy in determining properties of neutron stars for a much
wider class of gravity models.

\section{Acknowledgements}
This work is supported in part by the National Natural Science
Foundation of China under Grant No. 11275098, 11275067 and
11320101004, the State Scholarship Fund of China Scholarship Council
(CSC), the US National Science Foundation under Grant No.
PHY-1068022 and the CUSTIPEN (China-U.S. Theory Institute for
Physics with Exotic Nuclei) under DOE grant number
DE-FG02-13ER42025.

\bibliography{ReferencesFJF}
\vfill\eject
\end{document}